\documentstyle[epsfig]{aipproc}

\begin{document}
\title{Application of the synchrotron proton
blazar model to BL Lac objects}

\author{Raymond J. Protheroe$^*$ and Anita M\"ucke$^{\dagger}$}
\address{$^*$Department of Physics and Mathematical Physics,\\
The University of Adelaide, SA 5005, Australia\\
$^{\dagger}$Universit\'e de Montr\'eal, D\'epartement de Physique,
Montr\'eal, H3C 3J7, Canada}

\maketitle

\begin{abstract}
We apply the synchrotron proton blazar (SPB) model to the April
1997 flare of Markarian 501 and find we are able to fit the
observed spectral energy distribution.  We explore the effect of
target photon density on the high energy part of the spectral
energy distribution (SED) for fixed assumed magnetic field,
emission region size and Doppler factor and find that the 
luminosity and peak frequency of the high energy part of the SED may depend on
the
luminosity of the low energy part of the SED in high-frequency peaked
BL Lac objects (HBL).
\end{abstract}

\section*{Introduction}

The spectral energy distribution of blazars typically has a
double-humped appearance usually interpreted in terms of
synchrotron self-Compton models.  In proton blazar models, the
SED is instead explained in terms of acceleration of protons and
subsequent cascading.  We discuss a variation of the Synchrotron
Proton Blazar model in which the low energy part of the SED is
mainly synchrotron radiation by electrons ($e$) co-accelerated with
protons ($p$) which are assumed responsible for the high energy part of
the SED.  We consider the case where the maximum energy of the
accelerated protons is above the threshold for pion
photoproduction interactions on the synchrotron photons of the
low energy part of the SED.  \\

Using a Monte Carlo/numerical technique to simulate the
interactions and subsequent cascading of the accelerated protons,
we are able to fit the observed SED of the HBL
Markarian 501 during the April 1997 flare for a 12
hour variability time scale.  The parameters used for modeling
the April 1997 flare are: $D = 12$, $B\approx 20$~G, radius of
the emission region $R_{\rm{blob}}=8 \times 10^{15}$~cm, giving
a photon energy density of this radiation field of
$u_{\rm{target}} = 60$~GeV/cm$^{-3}$.  
The relevant radiation and loss time scales for photomeson
production, Bethe-Heitler pair production, $p$ synchrotron
radiation, and adiabatic losses due to jet expansion, are shown
in Fig.~1 together with the acceleration time scale. Proton synchrotron
losses, which turn out to be at least as important as losses due
to photopion production in our model, limit the injected $p$
spectrum to a Lorentz factor of $\gamma_p \approx 3 \times
10^{10}$ for the assumed model parameters.  We adopt a Kolmogorov
spectrum of turbulence for the magnetic field structure, and from
variability arguments constrain the shock angle (angle between
magnetic field and shock normal) to $\theta_1 \geq 75^\circ$.
Note that due to the non-zero shock angle, the acceleration time
scale shown in Fig.~1 does not follow a strict power-law, but is
curved (see ref.~\cite{MueckeProtheroe2000} for details).

The accelerated $p$
are assumed to follow a power law $\propto \gamma_p^{-2}$ between
$2\leq \gamma_p \leq \gamma_{p,\rm{max}} = 3\times 10^{10}$, and
in order to fit the emerging cascade spectra to the data a $p$
number density of $n_p \approx 250$~cm$^{-3}$, corresponding to
an energy density of accelerated protons of $u_p \approx
11.6$~TeV/cm$^{-3}$ is required.  With a magnetic field energy
density of $u_B \approx 11.7$~TeV/cm$^{-3}$ our model satisfies
$u_{\rm{target}} \leq u_p \approx u_B$ (all parameters are in the
co-moving frame of the jet), confirming that a significant
contribution from inverse-Compton scattering is not expected.

\begin{figure}[h] 
\centerline{\epsfig{file=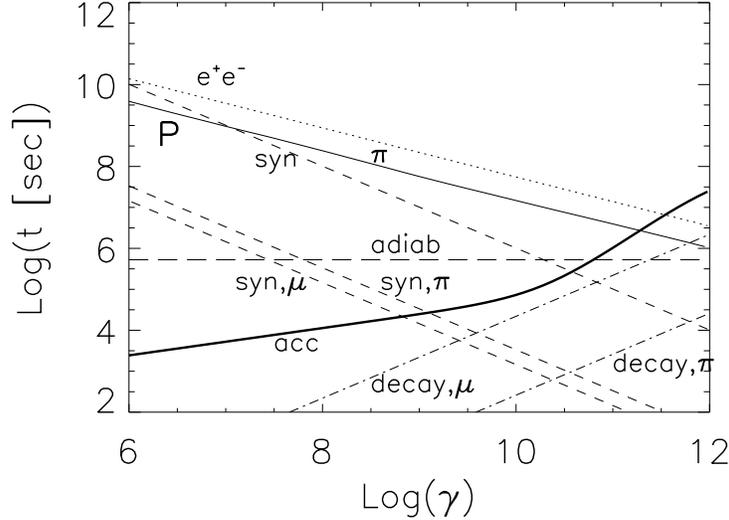,width=4.5in}}
\vspace{10pt}
\caption{ Mean energy loss time of $p$ for $\pi$-photoproduction
($\pi$), Bethe-Heitler pair production ($e^+e^-$) and synchrotron
radiation (syn).  Loss times for $\pi^\pm$- and $\mu^\pm$ for
synchrotron radiation (syn $\pi$, syn $\mu$) are also shown and
compared with their mean decay time scales (decay $\pi$, decay
$\mu$). The acceleration time scale (acc), based on Kolmogorov
turbulence, is calculated for a compression ratio of 4, a shock
velocity of $0.5c$ and shock angle $\theta_1 = 85^\circ$. The
adiabatic loss time (adiab) is assumed to be $2|B/\dot B| \approx
R/u_1 \approx D t_{\rm{var}}$.  We adopt $B\approx 20$~G, and all
quantities are in the co-moving frame of the jet.
(From ref.~\protect\cite{MueckeProtheroe2000})}
\label{fig2}
\end{figure}

\section*{Results and Discussion}

We use the
Monte-Carlo technique to simulate particle production and cascade
development, and this allows us to use exact cross sections.  For
photomeson production we use the Monte-Carlo code SOPHIA
\cite{Muecke99}, and Bethe-Heitler pair production is simulated
using the code of Protheroe \& Johnson
\cite{ProtheroeJohnson96}. We calculate the yields for both
processes separately, and the results are then combined according
to their relative interaction rates.
A detailed description of the
code is given in ref.~\protect\cite{MueckeProtheroe2000}. 

The results for the Markarian 501 flare are shown in Fig.~2.
We find that the emerging cascade spectra
initiated by gamma-rays from $\pi^0$ decay and by $e^\pm$ from
$\mu^\pm$ decay turn out to be relatively featureless,
synchrotron radiation produced by $\mu^\pm$ from $\pi^\pm$ decay,
and even more importantly by protons, and subsequent
synchrotron-pair cascading, is able to reproduce well the high
energy part of the SED.  For this fit we find that synchrotron
radiation by protons dominates the TeV emission, pion
photoproduction being less important with the consequence that we
predict a significantly lower neutrino flux than in other proton blazar models.

\begin{figure}[h] 
\centerline{\epsfig{file=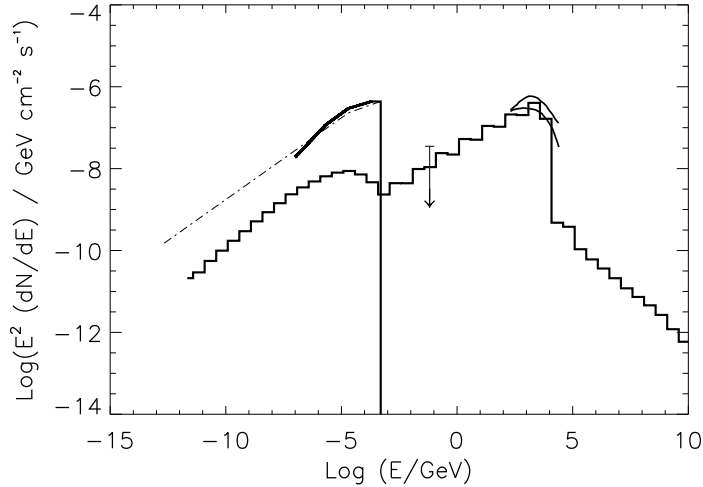,width=4in}}
\vspace{10pt}
\caption{ Best-fit model (histogram) in comparison with the data
of the 16 April 1997-flare of Mkn~501. Photon absorption on the
cosmic diffuse background radiation field is not
included. Straight solid lines: parametrization of the observed
synchrotron spectrum (BeppoSAX \& OSSE) and observed TeV-emission
corrected for cosmic background absorption
for two different IR background models \protect\cite{BednarekProtheroe99}; the
100~MeV upper limit is from ref.~\protect\cite{Catanese97}
 (observed 9-15
April 1997);
 dashed-dotted line: input target spectrum. (From
ref.~\protect\cite{MueckeProtheroe2000}) }
 
\end{figure}

We are in the process of applying the SPB model to other blazars
and have made predictions of the high energy part of the SED at
the source (i.e. not taking account of propagation to Earth
through the IR background).  We show in Fig.~3 preliminary
results \cite{MueckeProtheroe2001} for HBLs assuming $B = 30$
Gauss, $D = 10$, and $R = 10^{16}$ cm, but for different
$e$-synchrotron luminosities.  The broken power law curves
at the left give the assumed $e$-synchrotron spectra 
(which serve as the target field for pair-cascading and
pion production) for
$\log(\nu L_\nu^{\rm max,syn}/$erg s$^{-1}$) = 42.5, 43.5, 44.5, and
45.5, and the histograms to the right give the corresponding
contribution to the SED due to $p$ acceleration and
interaction. The cutoff energy of the $p$ particle spectrum 
increases with decreasing target photon density $u_{\rm phot}$, and
is determined by pion production for high $u_{\rm phot}$ and by
$p$ synchrotron radiation for low $u_{\rm phot}$.

The required total jet luminosity is $(L_{\rm
jet}/10^{46}$erg s$^{-1}$) = 2, 2, 2 and 2.5 corresponding to
the four $\nu L_\nu^{\rm max,syn}$ values above. For
decreasing target photon densities (i.e. lower $\nu L_\nu^{\rm max,syn}$)
the contribution of the total power from accelerated $p$ to the low energy
part of
the SED diminishes due to the lower contribution of synchrotron
radiation from muons produced as a result of pion production
and as a result of lower target photon density for pair synchrotron
cascading. The ratio of the low to high energy peak of the cascade spectrum is
mainly determined by opacity effects.

\begin{figure}[h] 
\centerline{\epsfig{file=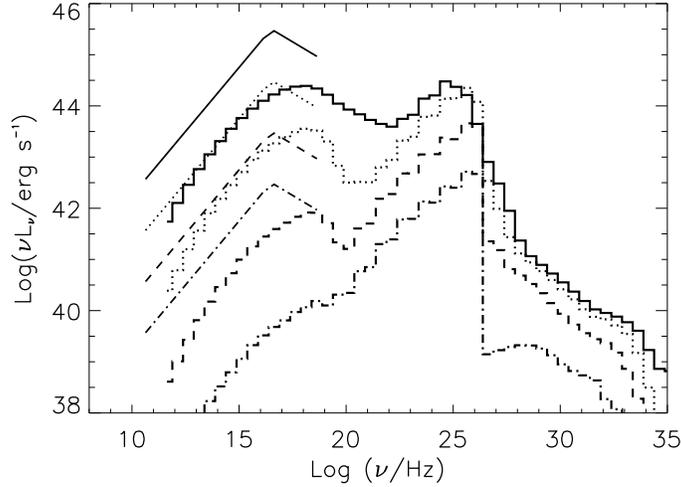,width=4in}}
\vspace{10pt}
\caption{SPB model predictions for HBL with broken power-law
synchrotron photon spectra $\propto \nu^{-1.5}$ below $\nu_b = 5\times
10^{16}$~Hz (observer frame) and $\propto \nu^{-2.25}$ above $\nu_b$ with
$\log(\nu L_\nu^{\rm max,syn}/$erg s$^{-1}$) = 42.5, 43.5, 44.5,
 45.5, and $B
= 30$ Gauss, $D = 10$, $u_B = u_P$, $R = 10^{16}$ cm (jet frame) giving
$\log(u_{\rm phot}$/eV cm$^{-3}$) = 9, 10, 11 and 12,
 respectively.}
\end{figure}

\end{document}